\documentclass{iopart}

\usepackage{graphicx}
\usepackage{amssymb}

\usepackage{natbib}
\newcommand{\uJy}{$\mu\mathrm{Jy}$ }

\def\newblock{\hskip .11em\@plus.33em\@minus.07em}

\newcommand{\aj}{{\it AJ} }

\newcommand{\Sa}{TVLM~513-46546~}
\newcommand{\SaS}{TVLM513~}

\newcommand{\SaPILim}{795~\uJy}
\newcommand{\SaPVLim}{743~\uJy}
\newcommand{\SaPulse}{2.4~mJy~}

\newcommand{\Sb}{2MASS~J0036+1821104~}
\newcommand{\SbS}{2M0036~}

\newcommand{\SbPILim}{942~\uJy}
\newcommand{\SbPVLim}{870~\uJy}

\newcommand{\SbLILim}{130~\uJy}
\newcommand{\SbLVLim}{95~\uJy}
\newcommand{\SbPulse}{2.8~mJy~}

\begin{document}

\title[Ultracool Dwarfs \SaS and \SbS]{325~MHz VLA Observations of Ultracool Dwarfs \Sa~and \Sb}
\author{T. R. Jaeger$^{1,2}$, R. A. Osten$^3$, T. J. Lazio$^4$, N. Kassim$^2$, and R. L. Mutel$^5$}
\address{$^1$ National Research Council Postdoctoral Research Associate}
\address{$^2$ US Naval Research Laboratory, Washington, DC 20375}
\address{$^3$ Space Telescope Science Institute, Baltimore, MD 21218}
\address{$^4$ Jet Propulsion Laboratory, Pasadena, CA 91109}
\address{$^5$ University of Iowa, Iowa City, IA 52242}
\ead{ted.jaeger.ctr@nrl.navy.mil}

\begin{abstract}
We present 325~MHz (90~cm wavelength) radio observations of ultracool dwarfs \Sa and \Sb using the Very Large Array (VLA) in June 2007. Ultracool dwarfs are expected to be undetectable at radio frequencies, yet observations at 8.5~GHz (3.5~cm) and 4.9~GHz (6~cm) of have revealed sources with $>$ 100~\uJy quiescent radio flux and $>$ 1~mJy pulses coincident with stellar rotation. The anomalous emission is likely a combination of gyrosynchrotron and cyclotron maser processes in a long-duration, large-scale magnetic field. Since the characteristic frequency for each process scales directly with the magnetic field magnitude, emission at lower frequencies may be detectable from regions with weaker field strength. We detect no significant radio emission at 325~MHz from \Sa or \Sb over multiple stellar rotations, establishing $2.5\sigma$ total flux limits of \SaPILim and \SbPILim respectively. Analysis of an archival VLA 1.4~GHz observation of \Sb from January 2005 also yields a non-detection at the level of $<$ \SbLILim. The combined radio observation history (0.3~GHz to 8.5~GHz) for these sources suggests a continuum emission spectrum for ultracool dwarfs which is either flat or inverted below 2-3~GHz. Further, if the cyclotron maser instability is responsible for the pulsed radio emission observed on some ultracool dwarfs, our low-frequency non-detections suggest that the active region responsible for the high-frequency bursts is confined within 2 stellar radii and driven by electron beams with energies less than 5~keV.
\end{abstract}

\submitto{\aj}
\maketitle

\section{Introduction}
Ultracool dwarfs (UCDs) describe a subsection of stellar objects located on the boundary between more massive stars and sub-stellar bodies such as gas giant planets. It includes fully convective, very low mass stars (M7 and lower) and all brown dwarfs. X-ray and H$\alpha$ intensity (typical proxies for magnetic activity) for UCDs is weak, dropping substantially after spectral class M7 \citep{Liebert1999ApJ,Audard2007AA}. Radio emission is expected to be undetectable at the \uJy level, based on empirical X-ray scaling laws \citep{Gudel1993ApJ,Benz1994AA}, and the assumption that highly neutral UCD atmospheres are incapable of sustaining magnetic stresses which pervade the atmospheres of solar-type stars. However, a growing number of UCDs have been discovered that display significant emission at cm-wavelengths and suggest the presence of persistent kG-scale magnetic fields \citep{Berger2001Nat,Burgasser2005ApJ,Hallinan2008ApJ,Osten2009ApJ}. Furthermore, the quiescent radio emission is nearly constant from spectral types M0 to L5 \citep{Berger2005ApJ,Berger2006ApJ}. It appears that, for at least some UCDs, the typical indicators of magnetic activity are not well correlated with decreased radio flux.

Radio emission from these peculiar UCDs is typically broadband and unpolarized with high brightness temperature ($10^8 - 10^9$K) during quiescence, and can exceed $10^{11}$K with nearly 100\% circular polarization during bursts \citep{Berger2002ApJ,Antonova2008AA,Hallinan2008ApJ}. There is also evidence of long-term radio variability \citep{Antonova2007AA}. The assumed radiation mechanism was initially incoherent gyrosynchrotron \citep{Berger2002ApJ,Berger2005ApJ} from populations of mildly relativistic electrons with a power-law energy distribution. However, the high brightness temperature and high circular polarization seen during burst events suggested a coherent radiation mechanism such as the cyclotron maser instability \citep[CMI,][]{Melrose1984JGR,Winglee1985JGR}. First suggested by \citet{Hallinan2006ApJ}, CMI has also been used to describe burst emission from the polar, low density - high magnetic field regions of magnetized planets \citep{Zarka1998JGR,Ergun2000ApJ}, Algol \citep{Mutel1998ApJ}, and late-type flare stars \citep{Bingham2001AA,Kellett2002MNRAS}. The CMI model may also explain the quiescent emission, possibly created via depolarization of persistent maser sources \citep{Hallinan2006ApJ,Littlefair2008MNRAS,Yu2011AA}. 

The exact mechanism(s) responsible for UCD radio emission (flaring on top of a quiescent background) is unclear, but it is likely due to a combination of both gyrosynchrotron emission and CMI. The emission frequency for each mechanism scales directly with the local magnetic field strength, emitting at the electron cyclotron frequency ($\Omega_{ce, MHz} = 2.8 \cdot B_G$). Gigahertz radio observations then require kG-scale fields, while detectable radio emission may exist at megahertz frequencies from regions of weaker field strength ($\sim$ 116~Gauss at 325~MHz). A vast majority of UCD observations have been limited to gigahertz frequencies where radio instruments are historically the most sensitive. Constraining the full spectral profile with addition of low frequency observations may provide valuable information to distinguish between the suggested emission processes, as well as reveal key differences in the atmospheres of UCDs compared to high mass M-dwarfs. Further, as UCDs are uniquely placed on the boundary between stars and sub-stellar objects, low frequency measurements also provide a guide for future observations of extrasolar planets.

\section{Target History}
\Sa (hereafter \SaS) is a spectral type M9 dwarf with mass equaling $0.09~M_{sun}$ and age $>$ 1~Gyr \citep{Reid2000AJ}. Radio emission from this near-by source ($\sim$ 10.6~pc, \citet{Dahl2002AJ}) was first detected by \citet{Berger2002ApJ}. Berger's observations at 8.5~GHz revealed both persistent stellar emission and occasional ($2\%-10\%$ duty cycle) pulses with flux densities exceeding 1~mJy. Detected bursts were highly circularly polarized and lasted multiple minutes. Both the persistent and burst emission features were later confirmed by \citet{Osten2006ApJ}, \citet{Hallinan2006ApJ}, \citet{Hallinan2007ApJ}, \citet{Berger2008ApJ}, \citet{Forbrich2009ApJ}, and \citet{Doyle2010AA} through observations at 8.5~GHz and 4.9~GHz. Further, \citet{Hallinan2007ApJ} and \citet[optical]{Lane2007ApJ} detected a pulse periodicity of $\sim$ 2 ~hr, consistent with the stellar rotation rate. \citet{Osten2006ApJ} made a 5$\sigma$ detection of \SaS at 1.4~GHz and it is undetected at lower frequencies at sensitivity levels similar to the GHz detections \citep{AntonovaThesis}.

\Sb (\SbS onward) is a L3.5 brown dwarf with mass $\sim 0.06 - 0.074~M_{sun}$ \citep{Schweitzer2001ApJ}, age $>$ 0.8~Gyr \citep{Burrows2001RevModPhys}, and is located at 8.8~pc \citep{Dahl2002AJ}. Radio observatons at 8.5~GHz by \citet{Berger2002ApJ} revealed quiescent emission with occasional bursts, similar to those witnessed on TVLM513. Subsequent measurements at 4.9~GHz by \citet{Hallinan2008ApJ} revealed strong (5x the quiescent level) circularly polarized pulses, lasting 5-20~min and with a 3~hr period corresponding to the stellar rotation rate. However, observations at 8.5~GHz by \citet{Berger2005ApJ,Forbrich2009ApJ} detected no significant activity over multiple stellar rotation periods, indicating a possible absence of periodicity above 4.9~GHz. To date, there are no published observations of \SbS below 4.9~GHz.

\section{Observations and Data Reduction}
\label{sec:obs}
Radio observations of UCDs \SaS and \SbS were conducted June $24-26$, 2007 using the NRAO Very Large Array (VLA). Each source was observed for $\sim$ 10.5~hr (10~s integration) using 2 x 6.25~MHz bands centered at 327.5~MHz and 321.6~MHz. Each frequency band was split into 15 spectral channels for the purpose of radio frequency interference removal and to mitigate bandwidth smearing. At observation time, the array utilized 23, 25~m antennas positioned in A configuration (maximum baseline $\sim$ 35~km), resulting in a $\sim$ 2.5~deg field of view and $\sim$ 6" x 5" angular resolution. The radio flux density scale was set using amplitude calibrator 3C~286 (assumed 24.49~Jy at 327.5~MHz). Phase calibration and the receiver bandpass correction was performed using standard sources 1513+236 and 0042+233 for \SaS and \SbS respectively. Measurements of each UCD and its corresponding phase calibrator were intertwined, performing 2 x 30~min primary target scans followed by a 5~min calibrator scan, then repeating.

Data reduction and imaging were performed using both AIPS \footnote{Astronomical Image Processing System, release 31DEC10} and Obit \footnote{Obit is developed and maintained by Bill Cotton at The National Radio Astronomy Observatory in Charlottesville, Virginia, USA and is made available under the GNU General Public License. version 1.1.269-6-64b. } software packages. The visibility data was calibrated using standard AIPS tasks. Automated data flagging, visibility self-calibration, and imaging was performed in Obit. Lightcurves of \SaS and \SbS were made in both total (Stokes I) and circularly polarized (Stokes V) flux using AIPS task DFTPL.

For comparison with published 1.4~GHz quiescent emission observations of TVLM513, we analyzed VLA archival data from January 10, 2005 which contained an $\sim$ 8~hr observation of 2M0036. The data consisted of 2 x 50~MHz bands centered on 1465~MHz and 1385~MHz obtained in the BnA hybrid configuration (maximum baseline $\sim$ 21.2~km NS and 12.2~km EW). Amplitude and phase offsets were determined using calibrators 3C~147 (assumed 21.39~Jy at 1465~MHz) and 0042+233. Data editing, calibration and imaging were performed using standard AIPS routines.

\section{Results}

\begin{figure}[h]
\begin{center}
\includegraphics[width=2.3in]{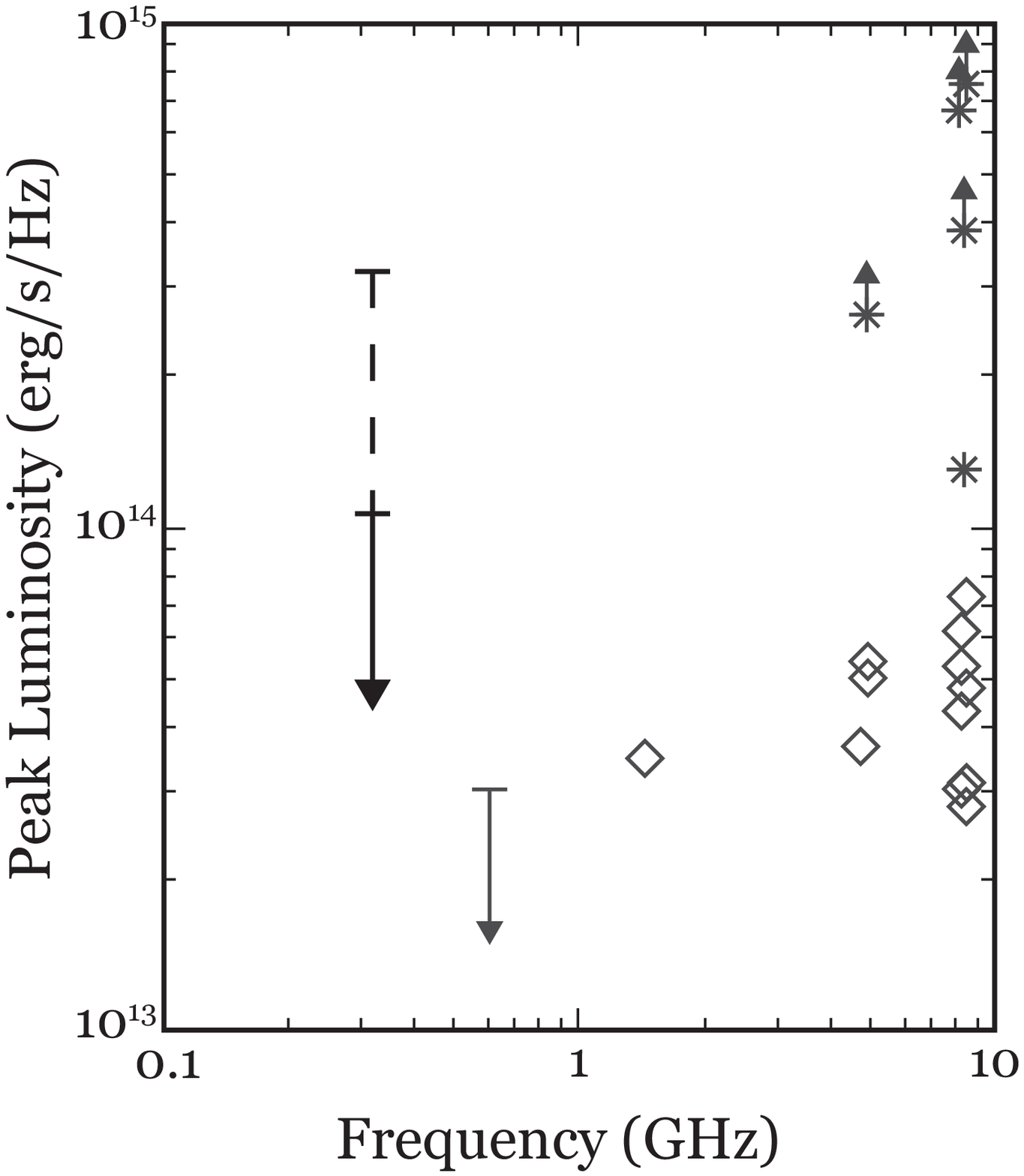}
\caption{Peak luminosity of \SaS from all published radio observations, scaling the radio flux density to a distance of 10.6~pc. Diamonds indicate measured quiescent emission and stars indicate burst/pulse emission. Reported error bars are roughly the size of each marker. Bursts reported as lower limits are displayed with upward arrows. Non-detection limits (quiescent solid, 10\% duty bursts dashed) are indicated with downward arrows ending at the 1$\sigma$ intensity. See Table \ref{ResultsTable} for specific radio measurements and author sources.}
\label{TVLM513Spec}
\ \\
\includegraphics[width=2.4in]{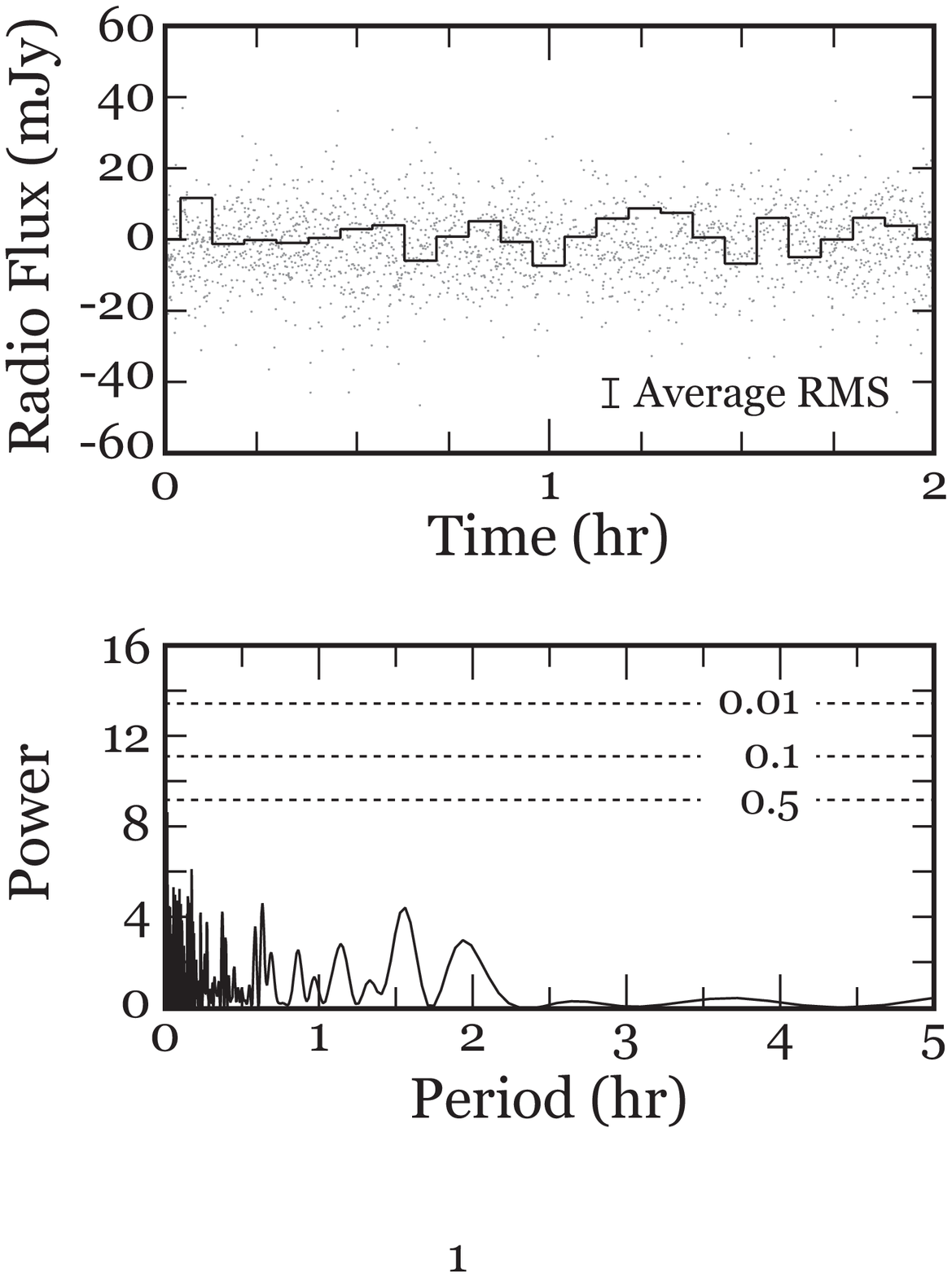}
\caption{(top) Measured circularly polarized radio flux at the position of \SaS folded with a period of 1.96~hr. Dots indicate the 10~s resolution measurements. The solid line indicates the 5~min median value of the folded data. (bottom) Lomb-Scargle periodogram of the 10~s resolution flux values. Dashed lines indicate false alarm probabilities of 0.01 (99\%), 0.1 (90\%), and 0.5 (61\%).}
\label{TVLM513LC}
\end{center}
\end{figure}

\begin{figure}[t!]
\begin{center}
\includegraphics[width=2.3in]{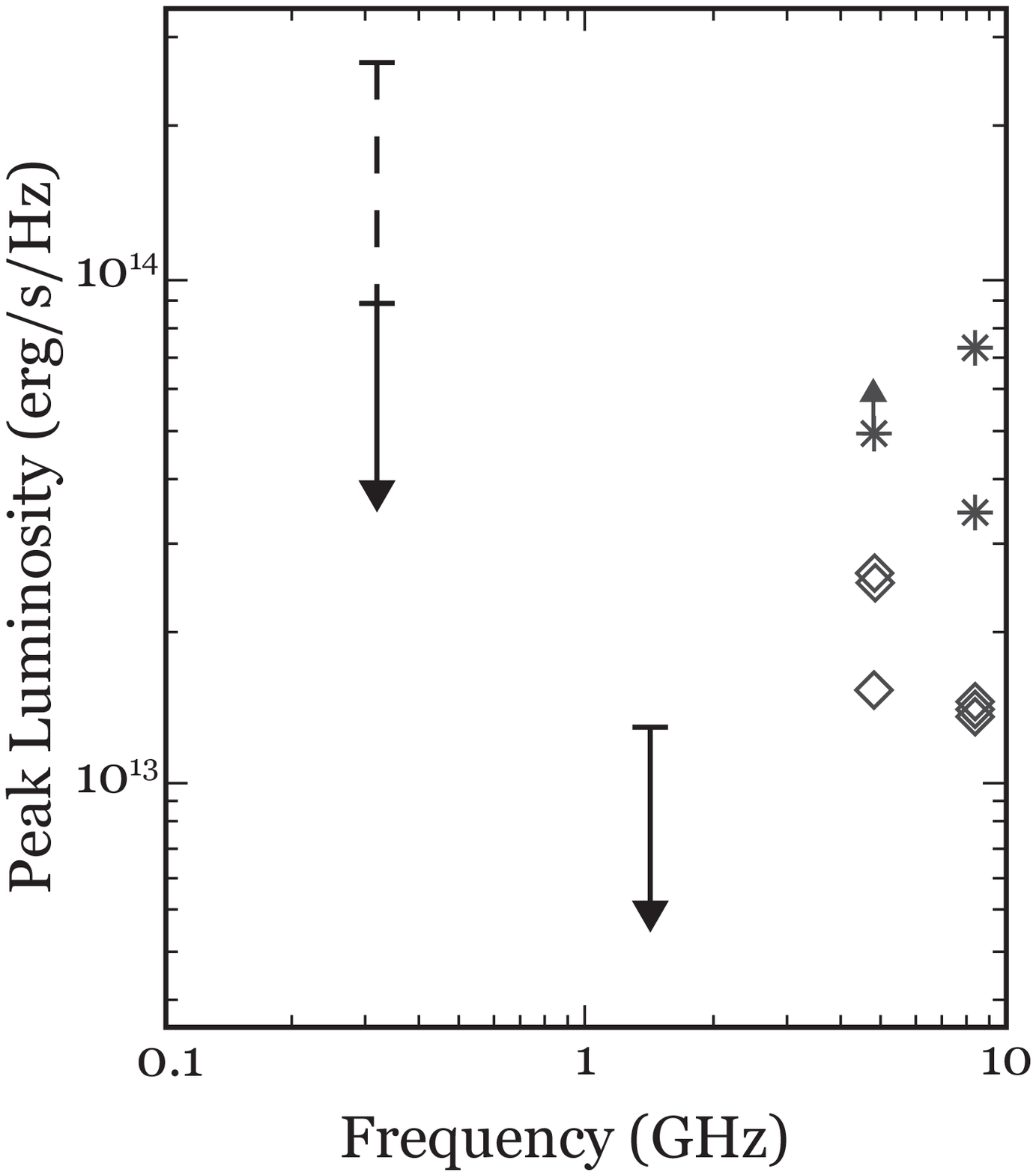}
\caption{Peak luminosity of \SbS from all published observations, scaling the measured radio flux density to a distance of 8.8~pc. Indicators are the same as in Figure \ref{TVLM513Spec}. See Table \ref{ResultsTable} for specific radio measurements and author sources.}
\label{2M0036Spec}
\ \\
\ \\
\ \\
\ \\
\includegraphics[width=2.4in]{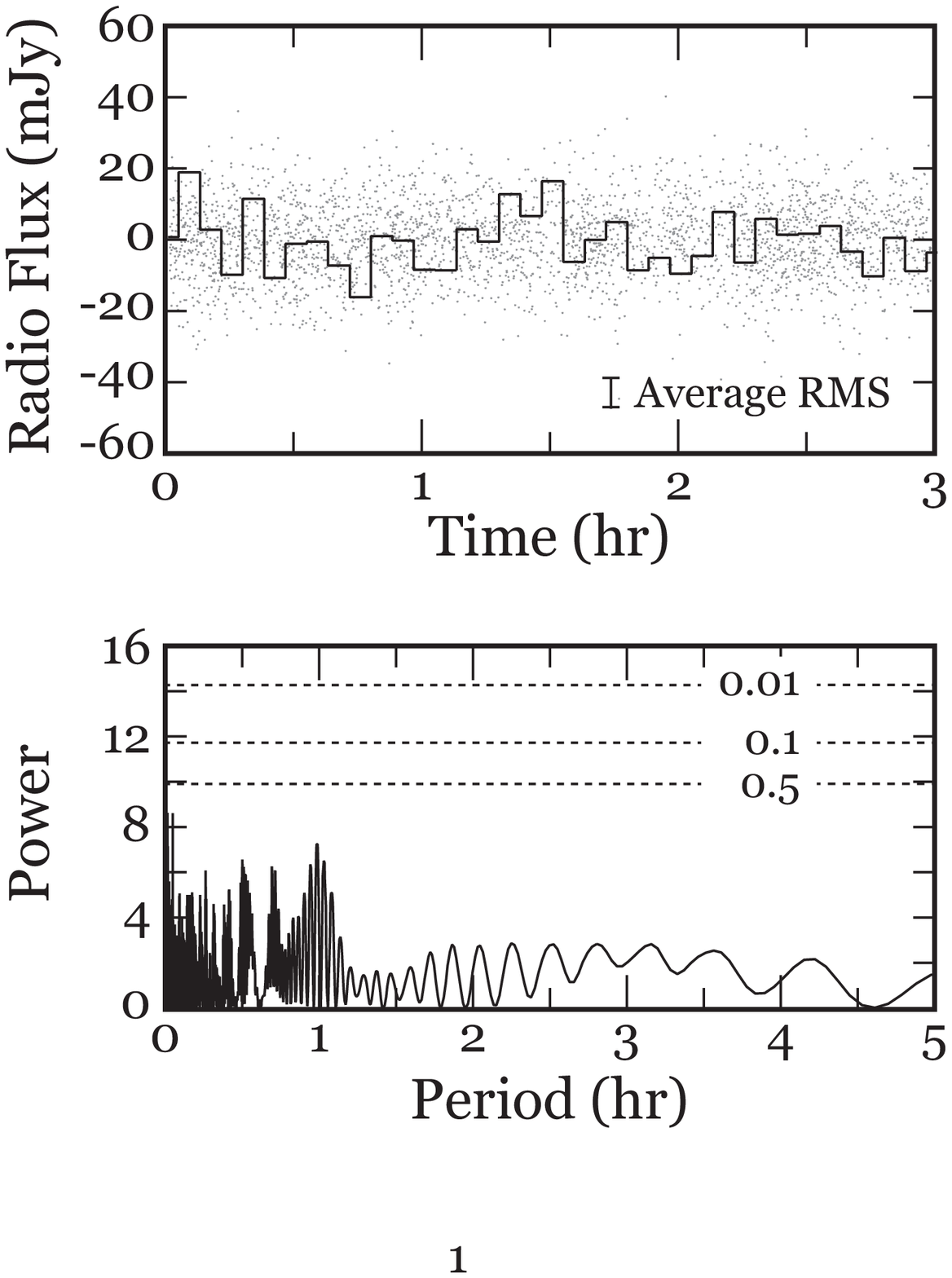}
\caption{(top) Measured radio flux (Stokes V) at the position of \SbS folded with a period of 3.08~hr. Dots represent the 10~s resolution radio intensity. A solid line indicates the 5~min median value of the 10~s measurements. (bottom) A Lomb-Scargle periodogram of the 10~s resolution circularly polarized flux values. Dashed lines indicate false alarm probabilities of 0.01 (99\%), 0.1 (90\%), and 0.5 (61\%).}
\label{2M0036LC}
\end{center}
\end{figure}

\subsection{\SaS}
We observe no significant unpolarized or circularly polarized radio emission associated with \SaS over the 10.5~hr observation.  The position of \SaS is well determined ($<$ 1"), based on closely spaced radio \citep{Berger2008ApJ} and infrared \citep{Cutri2003tmc} detections and estimates of TVLM513's  proper motion \citep{Schmidt2007AJ}. Given the positional accuracy, we report a $2.5\sigma$ non-detection limit to the total (Stokes I) and circularly polarized (Stokes V) quiescent radio flux. These limits are \SaPILim for I and \SaPVLim for V. The relationship between our $2.5\sigma$ upper limit for \SaS and previously measured peak flux density values is shown in Figure \ref{TVLM513Spec} (scaled to 10.6~pc) and listed in Table \ref{ResultsTable}. 

Previous observations of \SaS indicate a pulse period of $\sim$ 1.96~hr \citep{Hallinan2006ApJ,Lane2007ApJ}. To search for burst emission and any potential periodicity, we constructed lightcurves at the known position of \SaS in both unpolarized and circularly polarized intensity with time resolutions between 10~s and 10.5~hr.  We detect no significant variation to the measured radio intensity on any timescale. Also, no periodicity was found within the noise over $\sim$ 5 stellar rotations, performing both a blind period search and by folding the flux values at the expected period. Lightcurves of the circularly polarized flux with 10~s and 5~min temporal resolution, along with the corresponding Lomb-Scargle periodogram of the 10~s measurements are shown in Figure \ref{TVLM513LC}. Our 325~MHz non-detection implies a pulse flux density upper limit of \SaPulse (Stokes I) assuming a 10\% duty cycle. High frequency observations by \citet{Berger2002ApJ, Hallinan2007ApJ} observe pulse duty cycles in the range of 2-10\%. We choose a 10\% duty cycle to account for potential pulse dispersion at lower frequencies.    

\subsection{\SbS}
We searched the anticipated location of \SbS for polarized and unpolarized radio emission at observation frequencies of 325~MHz and 1.4~GHz. The position of \SbS is known to an accuracy smaller than each of the synthesized beams (see Sec. \ref{sec:obs}) and is therefore well constrained. No significant radio emission was observed at either frequency. Our 325~MHz non-detection establishes a $2.5\sigma$ upper limit on the the quiescent flux $<$ \SbPILim in total intensity (Stokes I) and $<$ \SbPVLim in circularly polarized (Stokes V) intensity for the 10.5~hr observation. The non-detection at 1400~MHz ($<$ \SbLILim in Stokes I, $<$ \SbLVLim Stokes V, $2.5\sigma$) sets an upper limit on the quiescent flux which is slightly lower than the predicted extrapolation from higher frequencies when assuming a flat spectrum (See Fig. \ref{2M0036Spec}).

No notable burst activity in the 325~MHz total flux or polarized flux measurements was observed on timescales from 10~s to 10.5~hr. The corresponding burst flux density upper limit is \SbPulse (Stokes I) assuming a 10\% duty cycle   We also detect no periodicity in the measured flux over $>$ 3 stellar rotations, assuming a stable rotation rate of 3.08~hr \citep{Hallinan2008ApJ}. A lightcurve of the 325~MHz circular polarization measurements folded at the expected pulse period and a Lomb-Scargle periodogram of 10~s resolution data are shown in Figure \ref{2M0036LC}. A similar flux variability search using the 1.4~GHz observation was not possible due to significant radio frequency interference which persisted for approximately 3~hrs during the session.

\begin{table}[t!]
\caption{Measurement summary listing observing frequency $\nu$ in GHz, observation length $\tau$ in hr, and the recorded flux S in \uJy. Measurements are made using the VLA unless noted otherwise.}
\begin{center}
\begin{tabular}{|c|c|c|c|l|l|l|}
\hline
Source	& Date		& $\nu$	& $\tau$    	  & S     		& Reference 				& Notes 	\\
\hline\hline
TVLM513	& 30.03.2008	& 8.5		& 7		  & $539\pm19$    	& \citet{Forbrich2009ApJ}	&		\\
		& 30.03.2008	& 8.5		& 7		  & $230\pm47$    	& 						& VLBA	\\
		& {\bf 26.07.2007} & {\bf 0.3} & {\bf 10} & {\bf $<$ 795} & {\bf This paper} & \\
		& 01.07.2007	& 8.4		& 8		  & $318\pm9$	    	& \citet{Doyle2010AA}		&		\\
		& 01.07.2007	& 8.4		& $-$	  & $>2900$		& 						& Burst	\\
		& 20.04.2007	& 8.5		& 9		  & $208\pm18$    	& \citet{Berger2008ApJ}		&		\\
		& 20.04.2007	& 8.5		& 8		  & $353\pm14$    	& 						&		\\
		& 20.04.2007	& 8.5		& $-$	  & $>5500$	    	&						& Burst	\\
		& 12.02.2007	& 0.6		& 4		  & $<225$	    	& \citet{AntonovaThesis}	& GMRT  \\ 
		& 20.05.2006	& 8.4		& 10		  & $464\pm9$	    	& \citet{Hallinan2007ApJ}	&		\\
		& 20.05.2006	& 8.4		& $-$	  & $> 5000$	    	& 						& Burst	\\
		& 20.05.2006	& 4.9		& 10		  & $368\pm16$    	& 						&		\\
		& 20.05.2006	& 4.9		& $-$	  & $> 2000$	    	& 						& Burst	\\
		& 13.01.2005	& 8.4		& 5		  & $396\pm16$    	& \citet{Hallinan2006ApJ}	&		\\
		& 13.01.2005	& 4.9		& 5		  & $405\pm18$    	& 						& 		\\
		& 24.01.2004	& 8.4		& 4		  & $228\pm11$    	& \citet{Osten2006ApJ}		&		\\
		& 24.01.2004	& 4.8		& 4		  & $284\pm13$    	& 						&		\\
		& 24.01.2004	& 1.4		& 4		  & $260\pm46$    	& 						&		\\
		& 23.09.2001	& 8.5		& $-$	  & $980\pm40$    	& \citet{Berger2002ApJ}		& Burst 	\\
\hline\hline
2M0036	& 30.03.2008	& 8.5		& 7		  & $144\pm22$	& \citet{Forbrich2009ApJ}	&		\\
		& {\bf 24.06.2007} & {\bf 0.3} & {\bf 10} & {\bf $<$ 942} & {\bf This paper} & \\
		& 24.09.2006	& 4.9		& 12		  & $241\pm14$	& \citet{Hallinan2008ApJ}	&		\\
		& 24.09.2006	& 4.9		& $-$	  & $>500$		& 						& Burst	\\
		& 10.01.2005	& 4.9		& 8		  & $152\pm9$		& \citet{Berger2005ApJ}		&		\\
		& {\bf 10.01.2005} & {\bf 1.4} & {\bf 8} & {\bf $<$ 130} & {\bf This paper} & \\
		& 28.09.2002	& 8.5		& 8		  & $134\pm16$	& \citet{Berger2005ApJ}		&		\\
		& 28.09.2002	& 4.9		& 8		  & $259\pm19$	& 						&		\\
		& 09.10.2001	& 8.5		& 3		  & $327\pm14$	& \citet{Berger2002ApJ}		& Burst	\\
		& 09.10.2001	& 8.5		& $-$	  & $720\pm40$	&						& Burst	\\
		& 23.09.2001	& 8.5		& 2	    	  & $135\pm14$	& 						&		\\
\hline
\end{tabular}
\end{center}
\label{ResultsTable}
\end{table}

\section{Discussion}
The absence of both flaring and quiescent emission from \SaS and \SbS implies constraints on the plasma environments (i.e. electron energy, spacial distribution, etc.) of UCDs. Of the two sources, \SaS may offer the most stringent constraints, given the numerous GHz observations of stable pulse activity in the year preceding our June 2007 measurements \citep[the closest burst is within 25 days,][]{Doyle2010AA}. Here, we explore what limits can be inferred by our non-detections within the context of behavior seen on active stars and magnetized planets.

\subsection{Flare Emission Constraints}
\citet{Hallinan2008ApJ} have argued that the primary mechanism responsible for time variable emission from UCDs is the cyclotron maser instability (CMI). The CMI mechanism requires (1) a continuous source of energetic beamed electrons accelerated by a parallel electric field and (2) a converging magnetic field topology. As the electron beam propagates toward increasing field strength, conservation of the first adiabatic invariant causes transfer of parallel to perpendicular energy. The resulting velocity distribution $f(v)$ becomes increasingly unstable ($\partial f/\partial v_ \bot  > 0$), leading to exponential wave growth at the electron cyclotron frequency. 

The CMI mechanism is quenched if the relativistic RX-mode cutoff frequency exceeds the cyclotron frequency. This constraint can be written in terms of an upper limit on the plasma $\beta$, the ratio of electron plasma to cyclotron frequencies \citep{Mutel2006JGR},
\begin{equation}
\beta = \frac{\omega_{pe}}{\Omega_{ce}} <  \sqrt{\frac{\gamma-1}{\gamma}},
\end{equation}
with $\gamma$ representing the electron beam Lorentz factor. For $\gamma-1\ll1$, this can be recast as 
\begin{equation}
\beta \sim 5\times10^{-3} \cdot \frac{n_e}{B_G^2\  E_{keV}} < 1
\end{equation}
where $n_e$ is the electron density per cubic centimeter, $B_G$ is the magnetic field in Gauss, and $E_{keV}$ is the beam energy in keV.

By assuming a magnetic field configuration, an electron density profile, and a mean beam energy, this constraint can be mapped to a coronal volume for which the CMI mechanism is viable. We assume a dipole field with surface equatorial field strength $B_o=5$ kG, consistent with estimates of 3 kG at the 8~GHz source altitude \citep{Reiners2007ApJ, Hallinan2008ApJ} but below the maximum of 10 kG based on dynamo models \citep{Browning2008ApJ}. The electron density profile is more problematic. Stellar coronal loop models suggest either a low-order power-law or exponential dependence with radial distance, with a scale height of order the loop size \citep{Rosner1978ApJ, Collier1988MNRAS}. We assume an example surface electron density $n_o = 10^9\ \rm{cm}^{-3}$ \citep{Fludra1999JGR,Yu2011AA} and an exponential dependence with a density scale height equal to the stellar radius.

\begin{figure}[ht]
\begin{center}
\includegraphics[width=5.25in]{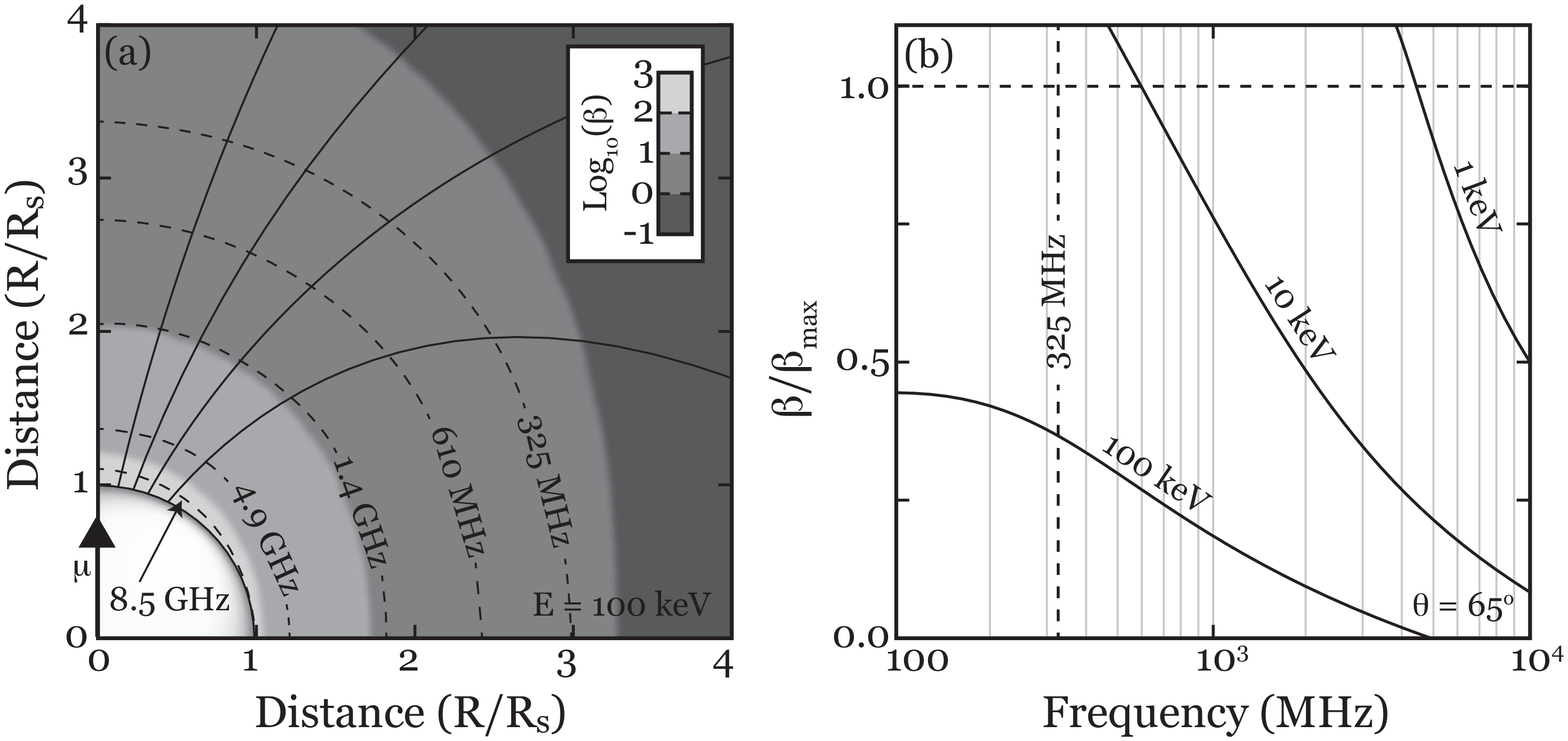}
\caption{(a) Masing criteria above the surface of a ultracool dwarf, assuming a dipole field ($B_o=5$~kG), an exponentially decreasing electron density ($n_o=10^9$~cm$^{-3}$), and an electron beam energy of 100~keV. Solid arcs are used to illustrate example magnetic field lines. Dashed curves indicate where the magnetic field magnitude is sufficient to produce cyclotron emission at 325~MHz, 610~MHz, 1.4~GHz, 4.9~GHz, and 8.5~GHz. White to black shading indicates the value of beta ($\beta = \omega_{pe}/\Omega_{ce}$). (b) Plasma beta versus electron cyclotron frequency at magnetic latitude 65$^o$ for electron beam energies of 1 keV, 10 keV, and 100 keV. Beta has been normalized to its maximum allowed value for positive CMI growth as given by equation (1). Note that at 325~MHz, only beam energies $E>10$ keV result in CMI growth ($\beta<1$), while above a few GHz, the CMI mechanism has positive growth at much lower energies.}
\label{SourceRegion}
\end{center}
\end{figure}

Figure~\ref{SourceRegion}a illustrates the masing criteria ($\beta$) in a model UDC magnetosphere for 100~keV electron beams. 
Presumably, the periodic bursts observed on UCDs are generated in an active region extending along an co-rotating field line and emitting at a frequency consistent with the local magnetic field strength. Detections at both 8.5~GHz and 4.9~GHz imply an active region extending to at least 1.5 stellar radii. This is consistent with estimates from \SbS by \citet{Hallinan2008ApJ}. Lower frequency observations probe larger radii, hence the highest observation frequency where $\beta<1$ and pulsed emission is absent defines the region's uppermost vertical extent. Assuming observations of \SbS and \SaS are typical of all UCDs and that the year-long history of pulse activity on \SaS continued throughout our 325~MHz observations (i.e. nearly 100\% pulse duty cycle), the absent pulse emission at 1.4~GHz \citep{Osten2006ApJ}, 610~MHz \citep{AntonovaThesis}, and 325~MHz (this work, for strong flat spectrum bursts) suggests that the pulse source is confined within 2 stellar radii.    

Figure~\ref{SourceRegion}b shows ratio of $\beta$ to the critical $\beta$ given in Equations 1-2 versus electron cyclotron frequency for a magnetic latitude of 65$^o$. For 325~MHz, the CMI mechanism is only viable for electron beam energies $E>20$ keV ($E>5$keV at 1.4~GHz), whereas only above a few GHz beam energies below 1~keV can drive the CMI mechanism. This may also account for our non-detection: If CMI is driven by relatively low-energy beams, it could account for CMI-induced bursts at high frequencies, but suppression at lower frequencies.


\subsection{Quiescent Emission Constraints}
The multi-frequency observations (325~MHz - 8.5~GHz) are consistent with a flat gyrosynchrotron spectrum with indication in the 1.4~GHz measurements of \SbS of a possible low-frequency spectral break. While our 325~MHz observational limits are the most stringent to date, they are not sensitive enough to imply any real constraints on the quiescent emission mechanism for UCDs, requiring at least a factor of 5-10 better sensitivity to rival the high frequency detections. The ability to make these needed measurements will rely on the completion of LOFAR (36 stations, $\sim$ 300~\uJy 2.5$\sigma$ at 250~MHz in 10~hr) or the extension of the EVLA Low-Band\footnote{The EVLA 74~MHz Low-Band system is scheduled to be completed Fall 2012} retrofit to 350~MHz ($\sim$ 100~\uJy 2.5$\sigma$ in 10~hr with 15\% fractional bandwidth).

\section{Summary and Conclusion}
We searched ultracool dwarfs \Sa and \Sb for radio emission at 325~MHz. Measurements of the total (Stokes I) and circularly polarized (Stokes V) flux were made on timescales between 10~s and 10.5~hr. While strong continuous emission and multi-minute, circularly polarized pulses were previously recorded at 4.9~GHz and 8.5~GHz, no significant emission was measured at 325~MHz. We set $2.5\sigma$ total flux limits of \SaPILim and \SbPILim for \Sa and \Sb, respectively, for the quiescent emission over 10.5~hr, consistent with a flat gyrosynchrotron spectrum with a potential spectral break near 2-3~GHz as indicated by \citep{Osten2006ApJ}. Furthermore, we observe no significant variation in the 325~MHz radio flux at the expected location of each source. The absence of variable radio flux from \Sa below 1.4~GHz suggests that pulse emission from these UCDs (assuming a CMI electron acceleration model) originates from a source region confined below 2 stellar radii and/or is driven by electron beam energies less than a few keV.        

\section{Acknowledgments}
This paper utilizes data from VLA programs AO218 and AB1169. We would like to thank Bill Cotton for assistance with Obit. The National Radio Astronomy Observatory is a facility of the National Science Foundation operated under cooperative agreement by Associated Universities, Inc. This research was performed while the primary author held a National Research Council Research Associateship Award a the US Naval Research Laboratory. Basic research in radio astronomy at the Naval Research Laboratory is supported by 6.1 base funding. Part of this research was carried out at the Jet Propulsion Laboratory, California Institute of Technology, under a contract with the National Aeronautics and Space Administration. The LUNAR consortium is funded by the NASA Lunar Science Institute to investigate concepts for astrophysical observatories on the Moon.

\newpage
\bibliography{StellarActivity}

\begin{thebibliography}{42}
\expandafter\ifx\csname natexlab\endcsname\relax\def\natexlab#1{#1}\fi
\expandafter\ifx\csname url\endcsname\relax
  \def\url#1{\texttt{#1}}\fi
\expandafter\ifx\csname urlprefix\endcsname\relax\def\urlprefix{URL }\fi
\providecommand{\eprint}[2][]{\url{#2}}

\bibitem[{{Antonova}(2007)}]{AntonovaThesis}
{Antonova}, A. 2007, Ph.D. thesis, The Queen's University of Belfast

\bibitem[{{Antonova} et~al.(2008){Antonova}, {Doyle}, {Hallinan}, {Bourke}, \&
  {Golden}}]{Antonova2008AA}
{Antonova}, A., {Doyle}, J.~G., {Hallinan}, G., {Bourke}, S., \& {Golden}, A.
  2008, \aap, 487, 317. \eprint{0805.4574}

\bibitem[{{Antonova} et~al.(2007){Antonova}, {Doyle}, {Hallinan}, {Golden}, \&
  {Koen}}]{Antonova2007AA}
{Antonova}, A., {Doyle}, J.~G., {Hallinan}, G., {Golden}, A., \& {Koen}, C.
  2007, \aap, 472, 257. \eprint{0707.0634}

\bibitem[{{Audard} et~al.(2007){Audard}, {Osten}, {Brown}, {Briggs},
  {G{\"u}del}, {Hodges-Kluck}, \& {Gizis}}]{Audard2007AA}
{Audard}, M., {Osten}, R.~A., {Brown}, A., {Briggs}, K.~R., {G{\"u}del}, M.,
  {Hodges-Kluck}, E., \& {Gizis}, J.~E. 2007, \aap, 471, L63.
  \eprint{0707.1882}

\bibitem[{{Benz} \& {Guedel}(1994)}]{Benz1994AA}
{Benz}, A.~O., \& {Guedel}, M. 1994, \aap, 285, 621

\bibitem[{{Berger}(2002)}]{Berger2002ApJ}
{Berger}, E. 2002, \apj, 572, 503. \eprint{arXiv:astro-ph/0111317}

\bibitem[{{Berger}(2006)}]{Berger2006ApJ}
--- 2006, \apj, 648, 629. \eprint{arXiv:astro-ph/0603176}

\bibitem[{{Berger} et~al.(2001){Berger}, {Ball}, {Becker}, {Clarke}, {Frail},
  {Fukuda}, {Hoffman}, {Mellon}, {Momjian}, {Murphy}, {Teng}, {Woodruff},
  {Zauderer}, \& {Zavala}}]{Berger2001Nat}
{Berger}, E., {Ball}, S., {Becker}, K.~M., {Clarke}, M., {Frail}, D.~A.,
  {Fukuda}, T.~A., {Hoffman}, I.~M., {Mellon}, R., {Momjian}, E., {Murphy},
  N.~W., {Teng}, S.~H., {Woodruff}, T., {Zauderer}, B.~A., \& {Zavala}, R.~T.
  2001, \nat, 410, 338. \eprint{arXiv:astro-ph/0102301}

\bibitem[{{Berger} et~al.(2008){Berger}, {Gizis}, {Giampapa}, {Rutledge},
  {Liebert}, {Mart{\'{\i}}n}, {Basri}, {Fleming}, {Johns-Krull}, {Phan-Bao}, \&
  {Sherry}}]{Berger2008ApJ}
{Berger}, E., {Gizis}, J.~E., {Giampapa}, M.~S., {Rutledge}, R.~E., {Liebert},
  J., {Mart{\'{\i}}n}, E., {Basri}, G., {Fleming}, T.~A., {Johns-Krull}, C.~M.,
  {Phan-Bao}, N., \& {Sherry}, W.~H. 2008, \apj, 673, 1080. \eprint{0708.1511}

\bibitem[{{Berger} et~al.(2005){Berger}, {Rutledge}, {Reid}, {Bildsten},
  {Gizis}, {Liebert}, {Mart{\'{\i}}n}, {Basri}, {Jayawardhana}, {Brandeker},
  {Fleming}, {Johns-Krull}, {Giampapa}, {Hawley}, \& {Schmitt}}]{Berger2005ApJ}
{Berger}, E., {Rutledge}, R.~E., {Reid}, I.~N., {Bildsten}, L., {Gizis}, J.~E.,
  {Liebert}, J., {Mart{\'{\i}}n}, E., {Basri}, G., {Jayawardhana}, R.,
  {Brandeker}, A., {Fleming}, T.~A., {Johns-Krull}, C.~M., {Giampapa}, M.~S.,
  {Hawley}, S.~L., \& {Schmitt}, J.~H.~M.~M. 2005, \apj, 627, 960.
  \eprint{arXiv:astro-ph/0502384}

\bibitem[{{Bingham} et~al.(2001){Bingham}, {Cairns}, \&
  {Kellett}}]{Bingham2001AA}
{Bingham}, R., {Cairns}, R.~A., \& {Kellett}, B.~J. 2001, \aap, 370, 1000

\bibitem[{{Browning}(2008)}]{Browning2008ApJ}
{Browning}, M.~K. 2008, \apj, 676, 1262. \eprint{0712.1603}

\bibitem[{{Burgasser} \& {Putman}(2005)}]{Burgasser2005ApJ}
{Burgasser}, A.~J., \& {Putman}, M.~E. 2005, \apj, 626, 486.
  \eprint{arXiv:astro-ph/0502365}

\bibitem[{Burrows et~al.(2001)Burrows, Hubbard, Lunine, \&
  Liebert}]{Burrows2001RevModPhys}
Burrows, A., Hubbard, W.~B., Lunine, J.~I., \& Liebert, J. 2001, Rev. Mod.
  Phys., 73, 719

\bibitem[{{Collier Cameron}(1988)}]{Collier1988MNRAS}
{Collier Cameron}, A. 1988, M.N.R.A.S., 233, 235

\bibitem[{{Cutri} et~al.(2003){Cutri}, {Skrutskie}, {van Dyk}, {Beichman},
  {Carpenter}, {Chester}, {Cambresy}, {Evans}, {Fowler}, {Gizis}, {Howard},
  {Huchra}, {Jarrett}, {Kopan}, {Kirkpatrick}, {Light}, {Marsh}, {McCallon},
  {Schneider}, {Stiening}, {Sykes}, {Weinberg}, {Wheaton}, {Wheelock}, \&
  {Zacarias}}]{Cutri2003tmc}
{Cutri}, R.~M., {Skrutskie}, M.~F., {van Dyk}, S., {Beichman}, C.~A.,
  {Carpenter}, J.~M., {Chester}, T., {Cambresy}, L., {Evans}, T., {Fowler}, J.,
  {Gizis}, J., {Howard}, E., {Huchra}, J., {Jarrett}, T., {Kopan}, E.~L.,
  {Kirkpatrick}, J.~D., {Light}, R.~M., {Marsh}, K.~A., {McCallon}, H.,
  {Schneider}, S., {Stiening}, R., {Sykes}, M., {Weinberg}, M., {Wheaton},
  W.~A., {Wheelock}, S., \& {Zacarias}, N. 2003, {2MASS All Sky Catalog of
  point sources.} (NASA/IPAC Infrared Science Archive)

\bibitem[{{Dahn} et~al.(2002){Dahn}, {Harris}, {Vrba}, {Guetter}, {Canzian},
  {Henden}, {Levine}, {Luginbuhl}, {Monet}, {Monet}, {Pier}, {Stone}, {Walker},
  {Burgasser}, {Gizis}, {Kirkpatrick}, {Liebert}, \& {Reid}}]{Dahl2002AJ}
{Dahn}, C.~C., {Harris}, H.~C., {Vrba}, F.~J., {Guetter}, H.~H., {Canzian}, B.,
  {Henden}, A.~A., {Levine}, S.~E., {Luginbuhl}, C.~B., {Monet}, A.~K.~B.,
  {Monet}, D.~G., {Pier}, J.~R., {Stone}, R.~C., {Walker}, R.~L., {Burgasser},
  A.~J., {Gizis}, J.~E., {Kirkpatrick}, J.~D., {Liebert}, J., \& {Reid}, I.~N.
  2002, \aj, 124, 1170. \eprint{arXiv:astro-ph/0205050}

\bibitem[{{Doyle} et~al.(2010){Doyle}, {Antonova}, {Marsh}, {Hallinan}, {Yu},
  \& {Golden}}]{Doyle2010AA}
{Doyle}, J.~G., {Antonova}, A., {Marsh}, M.~S., {Hallinan}, G., {Yu}, S., \&
  {Golden}, A. 2010, \aap, 524, A15+

\bibitem[{{Ergun} et~al.(2000){Ergun}, {Carlson}, {McFadden}, {Delory},
  {Strangeway}, \& {Pritchett}}]{Ergun2000ApJ}
{Ergun}, R.~E., {Carlson}, C.~W., {McFadden}, J.~P., {Delory}, G.~T.,
  {Strangeway}, R.~J., \& {Pritchett}, P.~L. 2000, \apj, 538, 456

\bibitem[{{Fludra} et~al.(1999){Fludra}, {Del Zanna}, {Alexander}, \&
  {Bromage}}]{Fludra1999JGR}
{Fludra}, A., {Del Zanna}, G., {Alexander}, D., \& {Bromage}, B.~J.~I. 1999,
  \jgr, 104, 9709

\bibitem[{{Forbrich} \& {Berger}(2009)}]{Forbrich2009ApJ}
{Forbrich}, J., \& {Berger}, E. 2009, \apjl, 706, L205. \eprint{0910.1349}

\bibitem[{{Guedel} \& {Benz}(1993)}]{Gudel1993ApJ}
{Guedel}, M., \& {Benz}, A.~O. 1993, \apjl, 405, L63

\bibitem[{{Hallinan} et~al.(2006){Hallinan}, {Antonova}, {Doyle}, {Bourke},
  {Brisken}, \& {Golden}}]{Hallinan2006ApJ}
{Hallinan}, G., {Antonova}, A., {Doyle}, J.~G., {Bourke}, S., {Brisken}, W.~F.,
  \& {Golden}, A. 2006, \apj, 653, 690. \eprint{arXiv:astro-ph/0608556}

\bibitem[{{Hallinan} et~al.(2008){Hallinan}, {Antonova}, {Doyle}, {Bourke},
  {Lane}, \& {Golden}}]{Hallinan2008ApJ}
{Hallinan}, G., {Antonova}, A., {Doyle}, J.~G., {Bourke}, S., {Lane}, C., \&
  {Golden}, A. 2008, \apj, 684, 644. \eprint{0805.4010}

\bibitem[{{Hallinan} et~al.(2007){Hallinan}, {Bourke}, {Lane}, {Antonova},
  {Zavala}, {Brisken}, {Boyle}, {Vrba}, {Doyle}, \& {Golden}}]{Hallinan2007ApJ}
{Hallinan}, G., {Bourke}, S., {Lane}, C., {Antonova}, A., {Zavala}, R.~T.,
  {Brisken}, W.~F., {Boyle}, R.~P., {Vrba}, F.~J., {Doyle}, J.~G., \& {Golden},
  A. 2007, \apjl, 663, L25. \eprint{0705.2054}

\bibitem[{{Kellett} et~al.(2002){Kellett}, {Bingham}, {Cairns}, \&
  {Tsikoudi}}]{Kellett2002MNRAS}
{Kellett}, B.~J., {Bingham}, R., {Cairns}, R.~A., \& {Tsikoudi}, V. 2002,
  \mnras, 329, 102

\bibitem[{{Lane} et~al.(2007){Lane}, {Hallinan}, {Zavala}, {Butler}, {Boyle},
  {Bourke}, {Antonova}, {Doyle}, {Vrba}, \& {Golden}}]{Lane2007ApJ}
{Lane}, C., {Hallinan}, G., {Zavala}, R.~T., {Butler}, R.~F., {Boyle}, R.~P.,
  {Bourke}, S., {Antonova}, A., {Doyle}, J.~G., {Vrba}, F.~J., \& {Golden}, A.
  2007, \apjl, 668, L163. \eprint{0709.1045}

\bibitem[{{Liebert} et~al.(1999){Liebert}, {Kirkpatrick}, {Reid}, \&
  {Fisher}}]{Liebert1999ApJ}
{Liebert}, J., {Kirkpatrick}, J.~D., {Reid}, I.~N., \& {Fisher}, M.~D. 1999,
  \apj, 519, 345

\bibitem[{{Littlefair} et~al.(2008){Littlefair}, {Dhillon}, {Marsh}, {Shahbaz},
  {Mart{\'{\i}}n}, \& {Copperwheat}}]{Littlefair2008MNRAS}
{Littlefair}, S.~P., {Dhillon}, V.~S., {Marsh}, T.~R., {Shahbaz}, T.,
  {Mart{\'{\i}}n}, E.~L., \& {Copperwheat}, C. 2008, \mnras, 391, L88.
  \eprint{0809.2193}

\bibitem[{{Melrose} et~al.(1984){Melrose}, {Dulk}, \&
  {Hewitt}}]{Melrose1984JGR}
{Melrose}, D.~B., {Dulk}, G.~A., \& {Hewitt}, R.~G. 1984, \jgr, 89, 897

\bibitem[{{Mutel} et~al.(2006){Mutel}, {Menietti}, {Christopher}, {Gurnett}, \&
  {Cook}}]{Mutel2006JGR}
{Mutel}, R.~L., {Menietti}, J.~D., {Christopher}, I.~W., {Gurnett}, D.~A., \&
  {Cook}, J.~M. 2006, Journal of Geophysical Research (Space Physics), 111,
  A10203. \eprint{arXiv:astro-ph/0609802}

\bibitem[{{Mutel} et~al.(1998){Mutel}, {Molnar}, {Waltman}, \&
  {Ghigo}}]{Mutel1998ApJ}
{Mutel}, R.~L., {Molnar}, L.~A., {Waltman}, E.~B., \& {Ghigo}, F.~D. 1998,
  \apj, 507, 371

\bibitem[{{Osten} et~al.(2006){Osten}, {Hawley}, {Bastian}, \&
  {Reid}}]{Osten2006ApJ}
{Osten}, R.~A., {Hawley}, S.~L., {Bastian}, T.~S., \& {Reid}, I.~N. 2006, \apj,
  637, 518. \eprint{arXiv:astro-ph/0509762}

\bibitem[{{Osten} et~al.(2009){Osten}, {Phan-Bao}, {Hawley}, {Reid}, \&
  {Ojha}}]{Osten2009ApJ}
{Osten}, R.~A., {Phan-Bao}, N., {Hawley}, S.~L., {Reid}, I.~N., \& {Ojha}, R.
  2009, \apj, 700, 1750. \eprint{0905.4197}

\bibitem[{{Reid} et~al.(2000){Reid}, {Kirkpatrick}, {Gizis}, {Dahn}, {Monet},
  {Williams}, {Liebert}, \& {Burgasser}}]{Reid2000AJ}
{Reid}, I.~N., {Kirkpatrick}, J.~D., {Gizis}, J.~E., {Dahn}, C.~C., {Monet},
  D.~G., {Williams}, R.~J., {Liebert}, J., \& {Burgasser}, A.~J. 2000, \aj,
  119, 369. \eprint{arXiv:astro-ph/9909336}

\bibitem[{{Reiners} \& {Basri}(2007)}]{Reiners2007ApJ}
{Reiners}, A., \& {Basri}, G. 2007, \apj, 656, 1121.
  \eprint{arXiv:astro-ph/0610365}

\bibitem[{{Rosner} et~al.(1978){Rosner}, {Tucker}, \& {Vaiana}}]{Rosner1978ApJ}
{Rosner}, R., {Tucker}, W.~H., \& {Vaiana}, G.~S. 1978, \apj, 220, 643

\bibitem[{{Schmidt} et~al.(2007){Schmidt}, {Cruz}, {Bongiorno}, {Liebert}, \&
  {Reid}}]{Schmidt2007AJ}
{Schmidt}, S.~J., {Cruz}, K.~L., {Bongiorno}, B.~J., {Liebert}, J., \& {Reid},
  I.~N. 2007, \aj, 133, 2258. \eprint{arXiv:astro-ph/0701055}

\bibitem[{{Schweitzer} et~al.(2001){Schweitzer}, {Gizis}, {Hauschildt},
  {Allard}, \& {Reid}}]{Schweitzer2001ApJ}
{Schweitzer}, A., {Gizis}, J.~E., {Hauschildt}, P.~H., {Allard}, F., \& {Reid},
  I.~N. 2001, \apj, 555, 368. \eprint{arXiv:astro-ph/0103402}

\bibitem[{{Winglee}(1985)}]{Winglee1985JGR}
{Winglee}, R.~M. 1985, \jgr, 90, 9663

\bibitem[{{Yu} et~al.(2011){Yu}, {Hallinan}, {Doyle}, {MacKinnon}, {Antonova},
  {Kuznetsov}, {Golden}, \& {Zhang}}]{Yu2011AA}
{Yu}, S., {Hallinan}, G., {Doyle}, J.~G., {MacKinnon}, A.~L., {Antonova}, A.,
  {Kuznetsov}, A., {Golden}, A., \& {Zhang}, Z.~H. 2011, \aap, 525, A39+.
  \eprint{1009.1548}

\bibitem[{{Zarka}(1998)}]{Zarka1998JGR}
{Zarka}, P. 1998, \jgr, 103, 20159

\end{thebibliography}

\end{document}